\documentclass[aps,prd,twocolumn]{revtex4}
\usepackage{epsf,epsfig,graphicx}
\begin{document}
\title{Comment on ``Nonuniversality of Transverse Momentum
Dependent Parton Distributions at Small x"}

\author{Hsiang-nan Li} 

\affiliation{Institute of Physics, Academia Sinica, Taipei, Taiwan
115, Republic of China}

\begin{abstract}
We point out that the analysis in arXiv:1003.0482 actually verifies
the universality of transverse momentum dependent quark
distributions at small $x$, which supports the observation in our
earlier work arXiv:0904.4150. Once the gluon exchanges responsible
for the additional infrared divergences are factorized and summed
into an exponential factor, it can be expressed as a matrix element
of Wilson line operators, and treated as a nonperturbative input to
the $k_T$ factorization of $pA$ collisions. The same exponential
factor has been extracted from $pp$ collisions in arXiv:0904.4150,
rendering possible experimental constraints on its behavior from
$pp$ processes, and predictions for $pA$ processes.
\end{abstract}


\maketitle

In a recent letter \cite{XY10}, Xiao and Yuan analyzed the initial
and final state interaction effects in dijet-correlations in $pA$
collisions at small $x$, and demonstrated the nonuniversality of the
transverse momentum dependent (TMD) quark distributions in a
scalar-QED model. The nonuniversality is caused by the infrared
divergence attributed to vanishing invariant mass squared
$-k_\perp^2$ of a Glauber gluon, $k_\perp$ being the gluon
transverse momentum. As a comparison, the all-order amplitudes for
the TMD quark distributions (see Eq.~(14) and Eq.~(39) in
\cite{XY102})
\begin{eqnarray}
\tilde A_{\rm DIS}&=&iV(r_\perp)\left[1-e^{-igg_1
W({\bf r}_\perp,{\bf R}_\perp)}\right],\nonumber\\
\tilde A&=&iV(r_\perp)\left\{1-e^{igg_1 [G({\bf R}_\perp+{\bf
r}_\perp)-G({\bf R}_\perp)]}\right\}\nonumber\\
& &\times e^{-igg_2G({\bf R}_\perp)},\label{1}
\end{eqnarray}
were extracted from the deep inelastic lepton-nucleus scattering
(DIS) and the dijet-correlation process $p+A\to Jet1+Jet2+X$,
respectively. The factor $G({\bf R}_\perp+{\bf r}_\perp)-G({\bf
R}_\perp)$ with $G({\bf R}_\perp)=K_0(\lambda R_\perp)/(2\pi)$,
$K_0$ being the Bessel function, and $\lambda$ an infrared
regulator, approaches $-W({\bf r}_\perp,{\bf
R}_\perp)=(1/2\pi)\ln(R_\perp/|{\bf R}_\perp+{\bf r}_\perp|)$ in the
$\lambda\to 0$ limit. The additional exponential factor
$e^{-igg_2G({\bf R}_\perp)}$ arises from the summation of the
Glauber divergences to all orders. The TMD quark distribution for
the nucleus $A$ from the dijet correlation was then given by
\begin{eqnarray}
\tilde q(x,q_\perp)&=&\frac{xP^{+2}}{8\pi^4}\int dp^- p^-\int
d^2R_\perp d^2R'_\perp d^2r_\perp\nonumber\\
& &\times e^{i{\bf q}_\perp\cdot({\bf R}_\perp-{\bf R}'_\perp)}
e^{-igg_2[G({\bf R}_\perp)-G({\bf R}'_\perp)]}
\nonumber\\
& &\times V(r_\perp)\{1-e^{igg_1[G({\bf R}_\perp+{\bf
r}_\perp)-G({\bf R}_\perp)]}\}
\nonumber\\
& &\times V(r'_\perp) \{1-e^{-igg_1[G({\bf R}'_\perp+{\bf
r}'_\perp)-G({\bf R}'_\perp)]}\},\label{2}
\end{eqnarray}
with ${\bf r}'_\perp={\bf R}_\perp+{\bf r}_\perp-{\bf R}'_\perp$. It
was claimed that the Glauber factor breaks the universality of the
TMD quark distribution \cite{XY10}.

We first point out that the same Glauber factor $e^{-iS({\bf b})}$
has been derived in our earlier work on the di-hadron production in
$pp$ collisions at small $x$ in the same model \cite{CLi09}, with
the exponent $S$ from the one-loop correction,
\begin{eqnarray}
S({\bf b})=\frac{g^2}{(2\pi)^2} \int
\frac{d^2k_\perp}{k_\perp^2+\lambda^2}e^{-i{\bf k}_\perp\cdot {\bf
b}}.\label{sf}
\end{eqnarray}
Equation~(\ref{sf}) leads to $G({\bf R}_\perp)$ in \cite{XY10} with
the impact parameter ${\bf b}$. The difference is that we did not
differentiate the coupling constants $g_2$ and $g$ associated with
the parton lines from the projectile and from the target,
respectively. The Feynman rule in Eq.~(\ref{sf}) was constructed by
means of the eikonal approximation for the Glauber gluons at small
$x$ \cite{CLi09}. Hence, it is straightforward to obtain the
definition of the Glauber factor
\begin{eqnarray}
e^{-igg_2G({\bf R}_\perp)}&=&\langle 0|W_-({\bf
R}_\perp;-\infty)^{\dag}
W_-({\bf R}_\perp;\infty)\nonumber\\
& &\times W_+({\bf 0};\infty)W_+({\bf
0};-\infty)^{\dag}|0\rangle,\label{soft}
\end{eqnarray}
where the Wilson line operator is written as
\begin{eqnarray}
\label{eq:WL.def} W_-({\bf R}_\perp;\infty) = P e^{-ig_2
\int_0^\infty dz u_-\cdot A({\bf R}_\perp+z u_-)}.
\end{eqnarray}
The expression of $W_+$ is similar, but with the light-cone vector
$u_-^\mu=(0,1,{\bf 0})$ being replaced by $u_+^\mu=(1,0,{\bf 0})$,
and $g_2$ by $g$. The net effect of $W_-({\bf
R}_\perp;-\infty)^{\dag}W_-({\bf R}_\perp;\infty)$ and $W_+({\bf
0};\infty)W_+({\bf 0};-\infty)^{\dag}$ demands the vanishing of the
components $k^+$ and $k^-$ of the gluon momentum, respectively. It
is easy to show, by expanding the Wilson line operators order by
order, that Eq.~(\ref{soft}) reproduces Eq.~(\ref{sf}).

Obviously, Eqs.~(\ref{1}) and (\ref{2}) indicate that the Glauber
divergences are factorizable in the impact parameter space. Once the
factorization is achieved, the Glauber factor can be treated as an
additional nonperturbative input to $pA$ collisions, just like the
soft factor introduced in the $k_T$ factorization \cite{CRS08}. The
same Glauber factor was derived in \cite{CLi09} and
\cite{XY10,XY102}, rendering possible experimental constraints on
its behavior from $pp$ processes, and predictions for $pA$
processes. Based on the observations that the Glauber factor can be
expressed as a matrix element of the Wilson line operators, treated
as a convolution piece in the $k_T$ factorization, and constrained
from some processes, we postulate that the analysis in \cite{XY10}
actually verifies the universality of TMD quark distributions at
small $x$ as claimed in \cite{CLi09}.

\end{document}